\documentclass[aps,pre,showpacs, twocolumn]{revtex4}

\usepackage{graphicx}
\usepackage{amssymb}
\usepackage{amsmath}
\usepackage{colordvi}
\usepackage{color}

\newcommand{\tbeta}{\tilde{\beta}}

\newcommand{\PT}{{\cal PT}}

\newcommand{\rev}[1]{\textcolor{black}{#1}}

\begin{document}

\title{Continuous families of non-Hermitian surface solitons}

\author{Dmitry A. Zezyulin}

\affiliation{Department of Physics and Engineering, ITMO University, St. Petersburg 197101, Russia}
\affiliation{Abrikosov Center for Theoretical Physics,  Moscow Institute of Physics and Technology, Dolgoprudny, Moscow Region 141701, Russia}

\date{\today}

\begin{abstract}

We show that surface solitons   form continuous families in one-dimensional  complex optical potentials of a  certain shape.  This result is   illustrated by  non-Hermitian gap-surface solitons at the interface between a uniform conservative medium and  a complex periodic potential. Surface soliton  families are parameterized by a real propagation constant.    The range  of possible propagation constants is  constrained    by the relation between   the continuous spectrum of the uniform medium and    the band-gap structure  of the   periodic potential. 

\end{abstract}

\maketitle

 Surface soliton is  a localized wave    propagating along  an interface between linear and nonlinear optical media   or between two different nonlinear    media. Early research in this field,  overviewed in e.g. \cite{Akhmediev,Boardman,Mihalache},    was mainly focused on surface modes at the boundary between two homogeneous media. Important developments have been made by  introducing  the concepts  of  discrete surface solitons \cite{Makris2005} and gap-surface solitons \cite{Karta2006}  at the interface between a layered (i.e., periodic) medium and a homogeneous  one. Surface solitons of these types have been studied thoroughly and observed in a  series of experiments, see reviews \cite{Lederer2008,Karta2009}. More recently, the steadily  growing interest in non-Hermitian physics has motivated intensive research of  gain-guided and dissipative surface waves and, in particular,  surface solitons supported by complex (i.e., non-Hermitian) optical potentials. It is well-known that  the  introduction of non-Hermiticity can  heavily impact the entire body of  solitons propagating in the system \cite{AA,Rosanov,KYZ}. In conservative systems,  solitary waves   usually form  continuous families which can be parameterized by a continuous change of a real propagation constant,     an energy flow,   or another   characteristics of the soliton. However, in a generic complex potential, the set of possible solitons  is typically    much scarcer:  instead of the continuous families, gain-guided solitary waves  most usually  exist   as isolated points, i.e., they can be found only at some discrete values of the propagation constant. Dynamical properties of such essentially dissipative solitons are   dramatically different from their conservative counterparts: in particular, stable gain-guided solitons dynamically behave as attractors and can be excited (by a nonresonant or resonant pump)  starting from a broad range of initial conditions that belong to the   basin of the  corresponding attractor. This is not the case of  conservative solitons whose propagation is constrained by the    energy flow conservation. In the meantime, there exist two    overlapping  classes of non-Hermitian potentials, where continuous families of solitons can exist. These classes   correspond to the  well-known $\PT$-symmetric potentials \cite{KYZ,Feng2017} and to the less studied   Wadati potentials which will be  discussed  below.  
 
 Essentially dissipative nonlinear surface modes and solitons have been studied   in a variety of previous publications, see e.g. \cite{Mihalache2008,Karta2010,Karta2012,Li2014,Karta2017, Chen2018,Dobrykh2018,Huang2019,Karta2019,KartaResonant2019}. It is remarkable that even  for   surface solitons at the interface of the truncated $\PT$-symmetric potential, the found solutions   still exist as isolated points   \cite{Karta2012}, because  the surface disrupts the global symmetry.  In the geometry with two transverse directions, families    of stable  surface modes have been found only in the case when the waveguide is $\PT$ symmetric   \emph{ along} the interface direction \cite{Li2014}.   The main goal of the present Letter  is to highlight   that there exists a broad class of one-dimensional   (1D)  complex optical potentials, where continuous families of surface solitons exist. 
 
 We model the propagation of the dimensionless light field  amplitude $\Psi(z,x)$ in the $z$-direction with the commonly used normalized nonlinear Schr\"odinger-type equation:
 \begin{equation}
 \label{eq:nls}
 i\Psi_z = -\Psi_{xx} - U(x)\Psi + g|\Psi|^2 \Psi. 
 \end{equation}
 In Eq.~(\ref{eq:nls}), complex-valued function $U(x)$ is the optical potential which describes weak modulation of the refractive index along the transverse $x$-direction.  Subscripts $z$ and $x$ mean derivatives with respect to the corresponding variables,    and $i$ is the imaginary unit.  Our study builds from the potentials of the form 
 \begin{equation}
 \label{eq:Wadati}
 U(x) = w^2(x) + iw_x(x),
 \end{equation}
 where    $w(x)$ is some real-valued differentiable function. We   call optical landscapes of the form (\ref{eq:Wadati})   \emph{Wadati potentials}, after the author of  Ref.~\cite{Wadati}, where the relevance of Eq.~(\ref{eq:Wadati})   was emphasized   in the context of $\PT$ symmetry. \rev{At the same time, it should   be noted that complex potentials of the form (\ref{eq:Wadati})    also appeared in much earlier literature. In particular, in the soliton theory potentials of this shape  were discussed as being closely related to the Miura transformation and to the Zakharov-Shabat spectral problem, see \cite{Newell}(Ch.~1) and\cite{Lamb}(Ch.~5).}

 For an even function $w(x)$ the potential $U(x)$ given by Eq.~(\ref{eq:Wadati}) is $\PT$ symmetric [recall that the standard definition of $\PT$-symmetry implies that $U(x) = U^*(-x)$, where the asterisk means complex conjugation].   At the same time, for a generic choice of function $w(x)$, the corresponding   potential is not $\PT$ symmetric. Realization of Wadati potentials for light propagating  in coherent atomic media was recently suggested in \cite{HHWadati}. It is    known that, when considered in the entire real axis $x\in \mathbb{R}$, Wadati potentials can support continuous families of nonlinear localized modes and solitons \cite{Tsoy,KonZez14,Yang20}.  This peculiarity   of    Wadati potentials  can be understood using the language of dynamical systems.   Let us briefly recall (and, at the same time, generalize) the arguments of Ref.~\cite{KonZez14}. We look  for  stationary modes in the form $\Psi(z,x) = e^{i\beta z} \psi(x)$, where $\beta$ is a real propagation constant, and  use the following representation: $\psi(x)=\rho(x) \exp\{i\int v(x) dx\}$, where $\rho(x)\geq 0$ and $v(x)$ are real-valued functions.   Using this substitution in Eq.~(\ref{eq:nls}) with a Wadati potential (\ref{eq:Wadati}), we obtain a system of coupled  differential equations for $\rho(x)$ and $v(x)$: 
 \begin{equation*}
 \rho_{xx} - \beta \rho + w^2\rho - g \rho^3 - v^2\rho=0, \quad
 2\rho_xv + \rho v_x + w_x\rho = 0.    
 \end{equation*}
 Treating these equations as  a dynamical system  where   $x$ plays  the role of an evolution  variable, we   find   an integral of motion:
 \begin{equation}
 \label{eq:integral}
 I = \rho_x^2 + \rho^2(v+w)^2 - \beta\rho^2 - g\rho^4/2, \quad dI/dx = 0.
 \end{equation}
 For a localized mode with   $\rho(x)$ and   $\rho_x(x)$ rapidly decaying to zero as $x\to\pm\infty$, this   quantity  must be  zero: $I=0$.

 We fix some  value of the propagation constant  $\beta$ and   assume that any solution $\psi_+(x)$ that tends to zero at $x\to+\infty$ [resp.,    $\psi_-(x)\to0$  at $x\to -\infty$] obeys  the asymptotic formula $\psi_+(x) = \Psi_+(x; \beta)(C_+ + o(1)_{x\to+\infty})$ [resp., $\psi_-(x) = \Psi_-(x; \beta)(C_- + o(1)_{x\to-\infty})$], where $\Psi_{\pm} = o(1)_{x\to\pm \infty}$ are   known functions, and $C_\pm$ are  arbitrary constants. Due to the phase-rotational invariance of the nonlinear Schr\"odinger equation, it is sufficient to consider  only real      $C_+$ and $C_-$.    To find a solution $\psi(x)$ which tends to zero both at $x\to+\infty$ and   $x\to-\infty$, it is necessary and sufficient to   find a pair $(C_+, C_-)$ such that the following three  equations hold:
 \begin{equation}
 \label{eq:3eqs}
 \rho_+(0) =  \rho_-(0), \quad  \rho_{+,x}(0) =  \rho_{-,x}(0), \quad v_+(0)=v_-(0).
 \end{equation}
 At   first glance,   system (\ref{eq:3eqs}) seems overdetermined. However,  the integral (\ref{eq:integral}) imposes  additional relations between  the functions: 
 \begin{equation}
 \rho_{\pm, x}^2(0) + \rho^2_\pm(0)(v(0)+w(0))^2 - \beta\rho_\pm^2(0) - g\rho_\pm^4(0)/2=0.
 \end{equation}
 These constraints imply that if any two equations (say the first two) of system (\ref{eq:3eqs}) are satisfied, then one of the following situations must take place:
 \begin{equation}
 v_+(0) = v_-(0) \quad \mbox{or} \quad v_+(0) + w(0) = -v_-(0) - w(0).
 \end{equation}
 The former case corresponds to a valid solution, and the latter case is a spurious solution which can be easily filtered out   in a practical realization of the numerical procedure.  Therefore, specifically     for just Wadati potentials,  system (\ref{eq:3eqs}) is not  overdetermined and can   have   a solution  or several solutions  that  correspond   to  physically meaningful solitons. Moreover, since the described procedure can be performed for any real propagation constant  $\beta$,   a   continuous family (or multiple families)  of localized modes can be constructed.
 
 The above considerations     have been  previously   used to find continuous families of nonlinear modes in asymmetric complex potentials   \cite{KonZez14}. Those families branched off  from linear eigenmodes   of the potential. However, it has not yet been appreciated that   Wadati potentials can also be used to create  a  medium that supports  families of surface modes. For instance, choosing a function $w(x)$ which is   constant for $x<0$ and periodic for $x>0$, one can find surface gap solitons which, in contrast to the analogous solutions in generic complex potentials,   exist as continuous families, rather than as isolated attractors. Moreover, the  above arguments can be easily extended on a more general class of potentials  composed of two Wadati functions:
 \begin{equation}
 \label{eq:concat}
 U(x) = \left\{
 \begin{array}{cc}
 w_1^2(x) + iw_{1,x}(x) \quad \mbox{for $x<0$},\\%
 w_2^2(x) + iw_{2,x}(x) \quad \mbox{for $x>0$},
 \end{array}
 \right.
 \end{equation}
 where   $w_1(x)$ and $w_2(x)$ satisfy the continuity condition:
 \begin{equation}
 \label{eq:cont}
 w_1(0) = w_2(0).
 \end{equation}
 Condition (\ref{eq:cont}) means that the real part  Re\,$U(x)$ is a continuous function, while the imaginary part Im\,$U(x)$ may have a jump at the interface, i.e., at $x=0$. Treating the semiaxes $x<0$ and $x>0$ separately, we find that in each semiaxis the system  has the integral of motion (\ref{eq:integral}), where $w$ should be replaced with $w_1$ or $w_2$. Then, in view of Eq.~(\ref{eq:cont}), any two equations of system (\ref{eq:3eqs}) again imply that the third equation automatically holds (up to a spurious solution mentioned above). The piecewise-defined potential in Eq.~(\ref{eq:concat}) can be  generalized to an arbitrary number of functions  $w_{1}(x), w_2(x), w_3(x)\ldots$ which are concatenated at arbitrary points $x_1<x_2<  \ldots$, provided that all functions satisfy the conditions $w_1(x_1) = w_2(x_1)$, $w_2(x_2) = w_3(x_2)$, etc.
 
 \begin{figure}
 	\begin{center}
 		\includegraphics[width=.998\columnwidth]{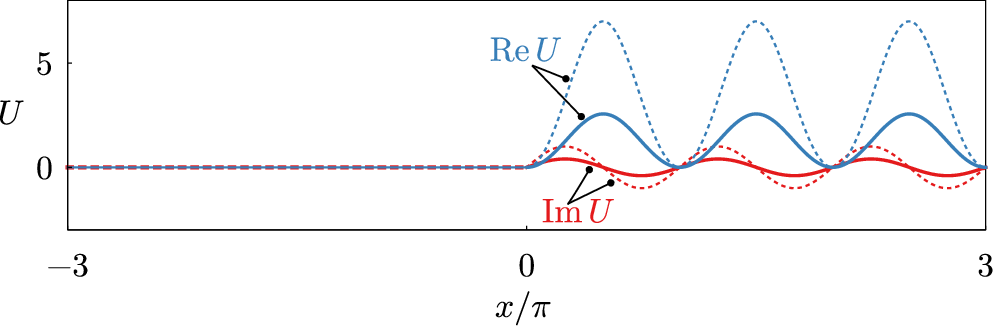}%
 		\caption{Real and imaginary parts of the potential  $U(x)$ chosen for the numerical illustration. Solid and dotted lines correspond to $\gamma=0.4$ and $\gamma=1$, respectively. Real parts are shown up to irrelevant additive constant (which can be absorbed by   a shift of the propagation constant). \label{fig:potential}}
 	\end{center}
 	
 \end{figure} 
 
 For a   numerical illustration, we focus on surface solitons at the interface of   a   complex periodic potential by choosing  the Wadati potential given by (\ref{eq:concat}) with $w_1(x) = \alpha/2$ and $w_2(x) = \alpha/2 + \gamma\sin^2 x$, where $\alpha$ is an auxiliary constant parameter that  tunes the  real background of the potential, and $\gamma>0$ is the amplitude of periodic modulation (see Fig.~\ref{fig:potential} for  plots  of the   potential).  The left semiaxis $x<0$ corresponds to a uniform  conservative medium. The  right halfspace $x>0$ is occupied by a complex-valued potential which (if considered on   the entire real axis) is   $\pi$-periodic and $\PT$ symmetric. The existence range of   surface solitons is constrained by the combination of two different requirements.  In the left semiaxis,  the  existence of a localized beam is possible only under the following requirement  $\tbeta:=\beta-\alpha^2/4>0$. If the latter inequality holds,  the asymptotic behavior of the decaying tail  can be easily found as  $\Psi_-(x; \beta) = e^{\sqrt{\tbeta } x}$. In the right semiaxis, a necessary requirement for the solution to be localized implies    that    $\tbeta$ belongs to a spectral gap of the   complex   periodic potential. By the Floquet theorem, in this case the asymptotic behavior is given as $\Psi_+(x; \beta) = e^{\lambda(\tbeta) x} u(x; \tbeta)$, where $\lambda(\tbeta)$ is the characteristic exponent (Re\,$\lambda<0$), and $u(x; \tbeta)$ is a $\pi$-periodic function of $x$. For each $\tbeta$ in the gap, $\lambda(\tbeta)$ and corresponding function $u(x; \tbeta)$ can be found numerically with the monodromy matrix approach. Using the established asymptotic behaviors, one can approximate the solutions for some      $x=\pm L$, where $L\gg 1$ is a sufficiently large number, and use the   Runge-Kutta method to compute  $(\rho_\pm(0), \rho_{x,\pm}(0), v_\pm(0))$ for any $C_+$ and $C_-$.

 \begin{figure}[t]
 	\begin{center} 	
 		\includegraphics[width=.998\columnwidth]{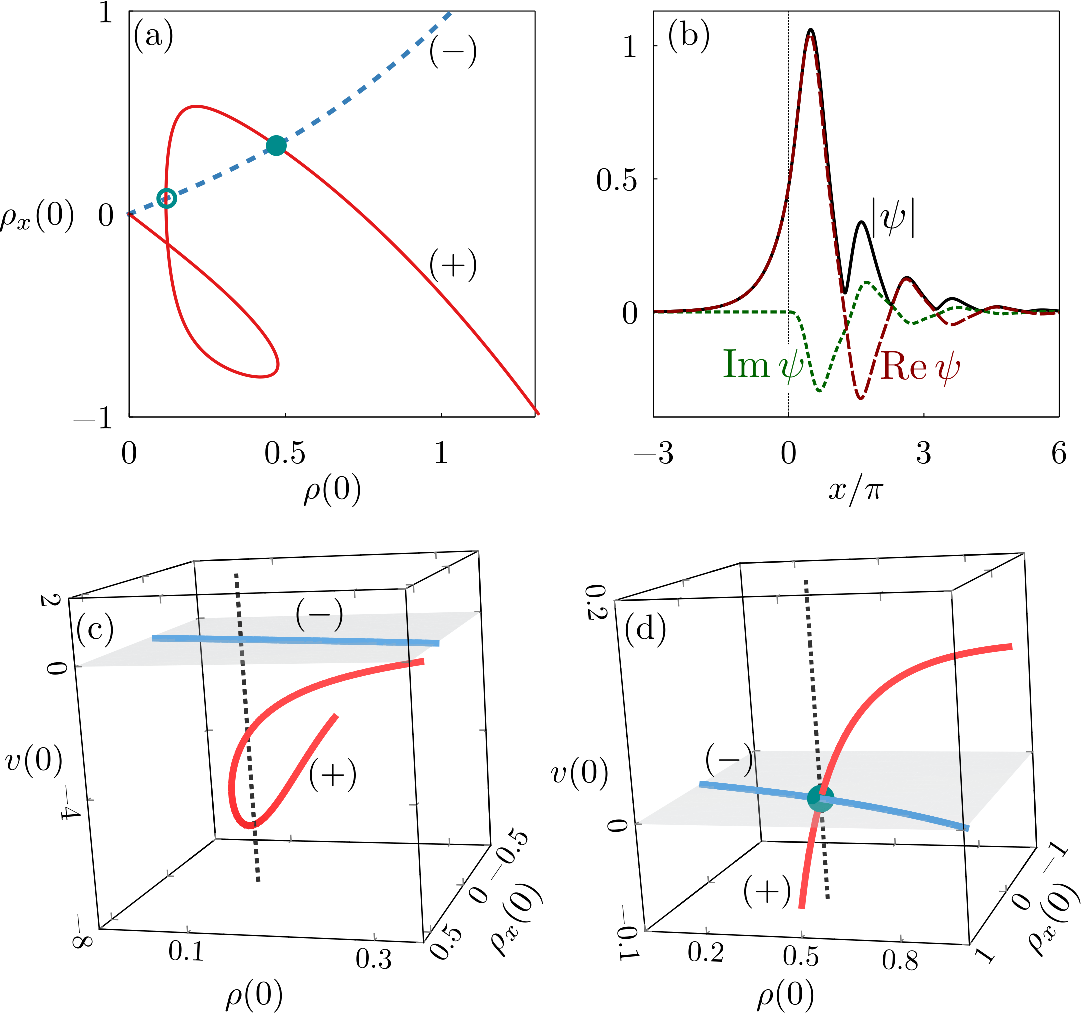}%
 		\caption{(a)  Dependencies $(\rho_\pm(0), \rho_{\pm, x}(0), v_\pm(0))$ plotted as 2D curves (labelled with ``$+$'' and ``$-$'', respectively) on the plane $(\rho(0), \rho_x(0))$. There are two intersections,  one of which   (marked with the open circle) corresponds  to a spurious solution, and another one (marked with the filled circle) is a valid solution corresponding to a surface soliton. Panels (c,d) show  the same dependencies in the 3D space (only the volumes close to the intersections are zoomed in): for the spurious   solution there is no intersection  in the 3D space (c), and  for the valid solution there is a    3D  intersection  (d).   Panel (b) illustrates the spatial profile  of the surface soliton obtained from   the valid intersection. \rev{Modulus of function $\psi$, its   real and imaginary parts are labelled as $|\psi|$, $\mathrm{Re}\,\psi$, and $\mathrm{Im}\,\psi$, respectively.} In this figure, $g=1$, $\alpha=6$, $\gamma = 0.4$, $\tbeta = 0.4$. \label{fig:diag}}
 	\end{center}
 \end{figure}

 To   illustrate the procedure of finding of   surface solitons, in Fig.~\ref{fig:diag}(a) we display a representative example of a diagram  obtained with   our  numerical  approach  for a particular value of the propagation constant $\tbeta=0.4$ and for coefficients $C_\pm$ varying within the following intervals: $C_+\in [0, 1.6]$, $C_-\in [0, 0.8]$. The 2D diagram   on the plane $(\rho(0), \rho_x(0))$   has two intersections, one of which corresponds to a spurious solution [no   intersection in 3D diagram plotted in Fig.~\ref{fig:diag}(c)], and  another one is indeed  a  valid solution that corresponds to a surface soliton [there is a  3D intersection in Fig.~\ref{fig:diag}(d)].  The surface soliton  obtained  from  the valid intersection is shown in Fig.~\ref{fig:diag}(b).  It    has complex internal structure with nontrivial real and imaginary parts. However, for $x\leq 0$, the imaginary part is identically zero, because   the corresponding halfspace is conservative.

 \begin{figure}[t!]
 	\begin{center}
 		\includegraphics[width=.998\columnwidth]{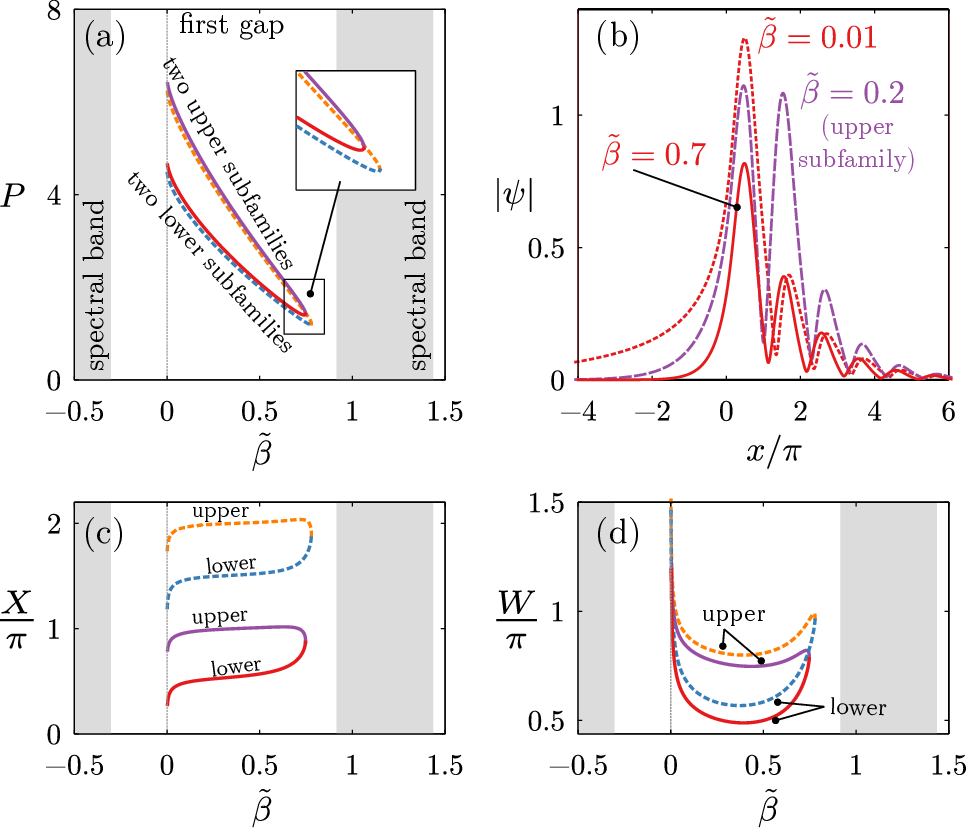}%
 		\caption{(a) Two families of surface solitons in the first finite gap plotted as the dependencies of the energy flow $P$  on   the shifted propagation constant $\tbeta$. Each family consists of an upper and a lower subfamily which   merge  at a   cutoff propagation constant (see the inset). Panels (c,d) display the   same (sub)families plotted as dependencies of the center-of-mass $X$ and meanwidth $W$ on $\tbeta$.  \rev{ The correspondence between different subfamilies in (a,c,d) can be established using the labels ``upper'' and ``lower'', solid and dashed lines, and colors (in the online version).} Panel (b) shows moduli  $|\psi|$ of two solitons from the lower subfamily closest to the interface at $\tbeta=0.7$ and $0.01$ and one soliton from the corresponding upper subfamily at $\tbeta=0.2$.  In this figure, $g=1$, $\alpha=6$, $\gamma = 0.4$. \label{fig:families}}
 	\end{center}
 	
 \end{figure}

 Solution plotted in Fig.~\ref{fig:diag}(b) belongs to a continuous family of surface solitons which can be obtained by varying the   propagation constant   and repeating the described procedure.  Moreover, we have found  that  extending the range of     parameters $C_+$ and $C_-$ it is possible to find multiple coexisting solutions and, respectively, multiple \rev{coexisting} soliton families.    Some of the found families, obtained in the defocusing medium for propagation constant lying in the first spectral gap,  are presented in  Fig.~\ref{fig:families}(a) as  the dependencies of the energy flow $P = \int_{-\infty}^\infty |\psi|^2dx$ on the propagation constant.  \rev{Each   family consists of two subfamilies (``upper'' and ``lower'') with   different values of $P$.}
 The found families have different right cutoff values where the upper and lower subfamilies merge   and disappear [see  the inset in Fig.~\ref{fig:families}(a)]. Thus, similar to surface gap solitons in a conservative medium \cite{Karta2006}, our solutions  exist only if the energy flow exceeds some nonzero threshold value.   The found solitons can be also characterized using  the center-of-mass $X = P^{-1} \int_{-\infty}^\infty x |\psi|^2 dx$  and meanwidth $W = \sqrt{  P^{-1} \int_{-\infty}^\infty (x-X)^2 |\psi|^2 dx}$,  plotted in Fig.~\ref{fig:families}(c,d). Figure~\ref{fig:families}(c) indicates  that there exists a sequence of families   consisting of solitons whose centers are situated at different distances from the interface position $x=0$. \rev{In Fig.~\ref{fig:families} we show  only two families (resp., four subfamilies) that consist of solitons centered closest to the surface; there exist other families with larger positive values of center-of-mass $X$. Solitons from these large-$X$ families are effectively situated in the bulk periodic medium and hence become similar to the conventional gap solitons. For this reason they  are not shown   in Fig.~\ref{fig:families}.}
 
 Surface solitons from lower subfamilies are   single-peaked,  while solitons from the upper subfamilies contain two out-of-phase peaks with close    amplitudes [see  an example with $\tbeta=0.2$  in Fig.~\ref{fig:families}(b)].  As the  propagation constant decreases, the surface solitons cease to exist in the limit $\tbeta\to +0$, where the left `halfsoliton' loses the localization, while the right tail of the soliton remains well-localized [see an example with $\tbeta=0.01$ in Fig.~\ref{fig:families}(b)]. As a result, in the limit $\tbeta\to +0$  the solitons centers incline downwards in Fig.~\ref{fig:families}(c).

 \begin{figure}[t!]
 	\begin{center}
 		\includegraphics[width=.998\columnwidth]{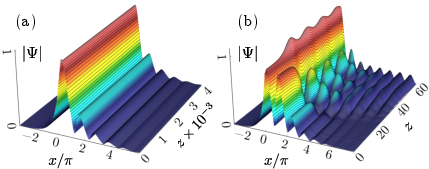}%
 		\caption{Propagation of stable (from the lower subfamily, panel  a) and unstable (from the upper subfamily, panel  b) surface solitons coexisting at $\tbeta=0.5$. Other parameters as in Fig.~\ref{fig:families}. \label{fig:dynamics} }
 	\end{center}
 	
 \end{figure} 
 
 Linear stability analysis and  dynamical simulations of soliton propagation indicate that solitons from     upper subfamilies are strongly unstable, while those from lower subfamilies are  stable.  Examples of stable and unstable propagations are presented in Fig.~\ref{fig:dynamics}. In this figure, each surface soliton has been initially perturbed by a $5\%$ complex-valued random noise and propagated according to Eq.~(\ref{eq:nls}). 
 
\begin{figure}
 	\begin{center}
 		\includegraphics[width=.998\columnwidth]{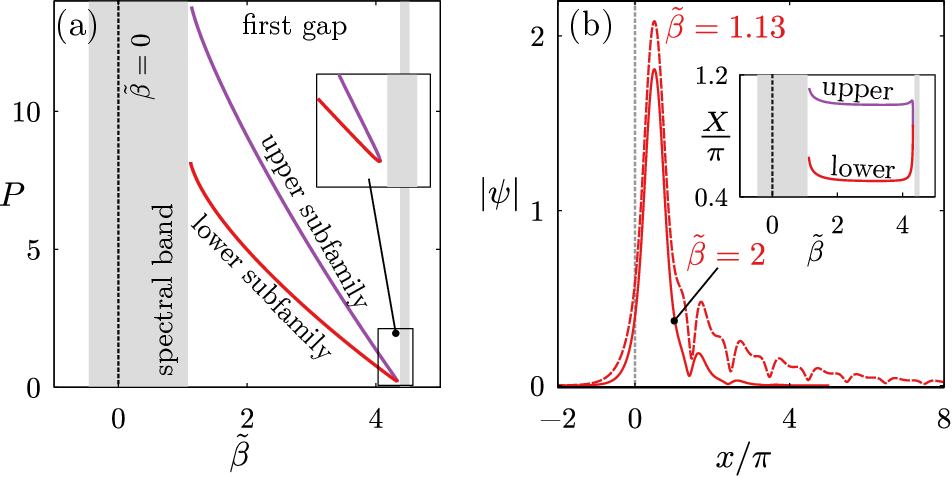}%
 		\caption{Families of surface solitons when the \textit{a priori} left existence boundary  $\tbeta=0$  lies inside a spectral band  (a) and examples of soliton profiles from the lower subfamily (b). Inset in (b) shows the centers-of-mass [compare with Fig.~\ref{fig:families}(c)].    In this figure, $g=1$, $\alpha=6$, $\gamma = 1$.  \label{fig:gamma=1}}
 	\end{center}
 \end{figure}

 Tuning the shape of the complex   lattice in the right halfspace,  it is possible to obtain a situation when the \emph{a priori}  left existence boundary   $\tbeta=0$   belongs to the spectral band (and not to the gap as in Fig.~\ref{fig:families}). In this case, the decrease of the propagation constant continues  the   soliton  families  up to the band edge, where the right halfsoliton becomes delocalized, while the left tail remains localized,  as illustrated in Fig.~\ref{fig:gamma=1}. 
 
 To conclude, we have demonstrated that non-Hermitian surface solitons   form continuous families in complex potentials of a  certain shape. This result has been illustrated for  gap-surface solitons guided by an interface between a homogeneous medium and a $\PT$-symmetric potential. The surface solitons have complex-valued internal structure and     feature different existence ranges and   delocalization scenarious depending on the relationship    between the edge of the continuous spectrum   of the uniform medium and the band-gap structure of the periodic potential. Regarding the future research, our results can be immediately generalized to surface solitons at an interface of a non-$\PT$-symmetric and/or nonperiodic potential. A generalization to a focusing medium is also straightforward.   Wadati potentials can be conveniently used to study surface soliton families in lattices with modulated separation between the cells (similar to those in  \cite{Karta2019}). A generalization  to two transverse directions is also possible.   An important byproduct  of our study is a   previously unexplored  class of \emph{layered} Wadati potentials composed of several    continuously concatenated  functions. These potentials   also admit    soliton families and are worth further  study.

\textbf{Funding.}   Ministry of Science and Higher Education of Russian Federation, goszadanie no. 2019-1246.

\end{document}